\documentstyle[12pt]{article}
\clearpage
\begin{document}

\topmargin 0pt
\oddsidemargin 7mm
\headheight 0pt
\topskip 0mm
\addtolength{\baselineskip}{0.40\baselineskip}

\hfill SOGANG-HEP 208/96

\hfill September 10, 1996

\vspace{0.5cm}

\begin{center}
{\large \bf BRST Quantization of the Proca Model
            based on the BFT and the BFV Formalism}
\end{center}

\vspace{0.5cm}

\begin{center}
Yong-Wan Kim$^{\dagger}$, Mu-In Park$^{\dagger}$, Young-Jai
Park$^{\dagger}$, and Sean J. Yoon$^{\dagger, *}$,  \\
\end{center}
\begin{center}
{\it \ $^{\dagger}$ Department of Physics and Basic Science Research
Institute \\
Sogang University, C.P.O. Box 1142, Seoul 100-611, Korea} \\
and\\
{\it \ $^{*}$ LG Electronics Research Center, Seoul 137-140, Korea}
\end{center}

\vspace{1cm}

\begin{center}
{\bf ABSTRACT}
\end{center}

The BRST quantization of the Abelian Proca model is performed using
the Batalin-Fradkin-Tyutin and the Batalin-Fradkin-Vilkovisky formalism.
First, the BFT Hamiltonian method is applied
in order to systematically convert a second class constraint
system of the model into an effectively first class one  by
introducing new fields.
In finding the involutive Hamiltonian we adopt a new approach which is more
simpler than the usual one.
We also show that in our model
the Dirac brackets of the phase space variables in the
original second class constraint system are exactly the same as
the Poisson brackets of
the corresponding modified fields in the extended phase space
due to the linear character of the constraints
comparing the Dirac or Faddeev-Jackiw formalisms.
Then, according to the BFV formalism we obtain that the desired resulting
Lagrangian preserving BRST symmetry in the standard local gauge fixing
procedure naturally includes the
St\"uckelberg scalar related to the explicit gauge symmetry breaking effect
due to the presence of the mass term.
We also analyze the nonstandard nonlocal gauge fixing procedure.

\vspace{1cm}

\newpage

\begin{center}
\section{\bf Introduction}
\end{center}

The Dirac method has been widely used in the Hamiltonian formalism
$^{1}$ to quantize the first and the
second class constraint systems generally, which do and do not form a closed
constraint algebra in Poisson brackets, respectively.
However, since the resulting Dirac brackets are generally field-dependent
and nonlocal, and have a serious ordering problem,
the quantization is under unfavorable circumstances
because of, essentially, the difficulty
in finding the canonically conjugate pairs.
On the other hand, the quantization of first class  constraint systems
established by Batalin, Fradkin, and Vilkovisky (BFV), $^{2,3}$
which does not have the previously noted problems of the Dirac method
from the start,
has been well appreciated in a gauge invariant manner with preserving
Becci-Rouet-Stora-Tyutin (BRST) symmetry. $^{4,5}$
After their works,
this procedure has been generalized to include the second class constraints
by Batalin, Fradkin, and Tyutin (BFT) $^{6,7}$
in the canonical formalism, and applied to various models
$^{8-10}$ obtaining the Wess-Zumino (WZ) actions. $^{11,12}$

Recently, Banerjee $^{13}$
has applied the BFT Hamiltonian method $^{7}$
to the second class constraint system of the Abelian Chern-Simons (CS)
field theory, $^{14-16}$ which yields the
{\it strongly involutive} first class
constraint algebra in an extended phase space
by introducing new fields.
As a result, he has obtained a new type of an Abelian WZ action,
which cannot be obtained in the usual path-integral framework.
Very recently, we have quantized several interesting models
$^{17}$ as well as the non-Abelian CS case, which yields the
{\it weakly involutive}
first class constraint system originating from
the non-Abelian nature of the second class constraints of the system,
by considering the generalized form of the BFT formalism. $^{18}$
As shown in all these works,
the nature of the second class constraint algebra originates from only the
symplectic structure of the CS term, not due to the local gauge symmetry
breaking.
Banerjee's and Ghosh $^{19}$
have also considered the Abelian and non-Abelian (incompletely) Proca model,
$^{20}$
which have the explicit gauge-symmetry breaking term by extending the BFT
approach to the case of rank-1 non-Abelian conversion compared to the Abelian
(rank-0) conversion.
As a result, the extra field in this approach has identified
with the St\"uckelberg scalar. $^{21}$
However, all these analysis do not carry out the covariant gauge fixing
procedure preserving the BRST symmetry based on the BFV formalism hence
completing the BRST quantization procedure.
Furthermore, up to now all above authors do not explicitly treat the Dirac
brackets
in this BFT formalism although the general but classical relation between
the Dirac bracket and the Poisson bracket in the extended phase space are
formally reported for the case of Abelian conversion. $^{7}$

In the present paper, the BRST quantization of the Abelian Proca model
$^{20}$ is performed completely by using the usual BFT $^{7}$ and the
BFV $^{2,3}$ formalism.
In section 2, we will apply the usual BFT formalism $^{7}$ to the Abelian
Proca model in order to convert the second class constraint system
into a first
class one by introducing new auxiliary fields.
Here, we newly obtain
the relation that the well-known Dirac brackets between the phase space
variables in our starting second class constraint system of the Abelian Proca
model are the same as the Poisson brackets of the corresponding modified ones
in the extended phase space without
$\Phi \rightarrow 0$ limiting procedure of the general formula of BFT $^{7}$
due to essentially the linear character
of the constraint. It is also compared with the Dirac
$^{1}$ or Faddeev-Jackiw (FJ) symplectic formalism, $^{22}$
which is to be regarded as the improved version of the Dirac one.
Furthermore, we adopt a new approach, which is more simpler than the usual
one, in finding the involutive Hamiltonian by using these modified
variables.
In section 3, we will consider the completion of the BRST quantization
based on the BFV formalism. As a result we show that
by identifying a new  auxiliary field with the
St\"uckelberg scalar we naturally derive the St\"uckelberg scalar
term related to the explicit gauge symmetry breaking mass term
through a BRST invariant and local gauge
fixing procedure according to the BFV formalism.
We also analyze the nonlocal gauge fixing procedure which has been recently
studied by several authors. $^{23}$
Our conclusions are given in section 4.

\vspace{1cm}
\begin{center}
\section {\bf BFT Formalism}
\end{center}

\begin{center}
\subsection {\bf Proca Model and  Constraints}
\end{center}

Now, we first apply the usual BFT formalism which assumes the Abelian
conversion of the second class constraint of the original system $^{7}$ to
the Abelian Proca model of the massive photon in four dimensions, $^{20}$
whose dynamics are given by
\begin{equation}
S = \int d^4x~
             [ -\frac{1}{4} F_{\mu\nu} F^{\mu\nu}
             + \frac{1}{2} m^2 A_\mu A^\mu ],
\end{equation}
where $F_{\mu\nu} = \partial_\mu A_\nu - \partial_\nu A_\mu$, and
$g_{\mu\nu} = diag(+,-,-,-)$.

The canonical momenta of gauge fields are given by
\begin{eqnarray}
\pi_{0} & \equiv & \frac{\delta S}{\delta \dot{A}_{0}} \approx 0, \nonumber\\
\pi_{i} & \equiv & \frac{\delta S}{\delta \dot{A}^{i}} = F_{i0}
\end{eqnarray}
with the Poisson algebra $\{ A^{\mu}(x), \pi_{\nu}(y) \}
=\delta ^{\mu}_{\nu} \delta({\bf x}-{\bf y})$.
The weak equality ` $\approx$ ' means the equality is not applied before
all involved calculations are finished. $^{1}$ In contrast, the strong
equality ` $=$ ' means the equality can be applied at all the steps of the
calculations.

Then, $\Omega_1 \equiv \pi_0 \approx 0$ is a primary constraint. $^{1}$
The total Hamiltonian is
\begin{equation}
H_{T}=H_{c}+\int d^{3}x u \Omega_{1}
\end{equation}
with the multiplier $u$ and the canonical Hamiltonian
\begin{equation}
H_c = \int d^3x \left[
                \frac{1}{2} \pi_i^2 + \frac{1}{4} F_{ij} F^{ij}
              + \frac{1}{2} m^2 \{ (A^0)^2 + (A^i)^2 \}
              - A_0\Omega_2
                                                              \right],
\end{equation}
where $\Omega_2$ is the Gauss' law constraint,
which comes from the time evolution of $\Omega_1$ with $H_{T}$, defined by
\begin{equation}
\Omega_2 = \partial^i \pi_i + m^2 A^0 \approx 0.
\end{equation}
Note that the time evolution of the Gauss' law constraint with $H_{T}$
generates no more
additional constraints but only determine the multipler
$ u \approx -\partial_{i}A^{i}$. As a result, the full constraints of this
model are
$\Omega_{i}~(i, j = 1, 2)$ which satisfy the second class constraint
algebra as follows
\begin{eqnarray}
&&\Delta_{ij}(x,y) \equiv \{ \Omega_i(x), \Omega_j(y) \}
               = - m^2 \epsilon_{ij} \delta^3({\bf x}-{\bf y}),
               \\  \nonumber
&& \mbox{det} \Delta_{ij}(x,y) \neq 0,
\end{eqnarray}
where we denote $x=(t,\bf{x})$ and three-space vector ${\bf x}=(x^1,x^2,x^3)$
and $\epsilon_{12}=-\epsilon_{21}=1$.

We now introduce new auxiliary fields $\Phi^i$ to convert the second
class constraints $\Omega_i$ into first class ones
in the extended phase space with the Poisson algebra
\begin{eqnarray}
   \{A^{\mu}(\mbox{or}~ \pi_{\mu}), \Phi^{i} \} &=& 0,  \nonumber \\
   \{ \Phi^i(x), \Phi^j(y) \} = \omega^{ij}(x,y) &=&
                      -\omega^{ji}(y,x).
\end{eqnarray}
Here, the constancy, i.e., the field independence of $\omega^{ij}(x,y)$, is
considered for simplicity.

According to the usual BFT method, $^{7}$
the modified constraints $\widetilde{\Omega}_{i}$ with the property
\begin{eqnarray}
\{\widetilde{\Omega}_{i}, \widetilde{\Omega}_{j} \}=0,
\end{eqnarray}
which is called the ${\it Abelian~ conversion}$, which is rank-0, of
the second
class constraint (6) are generally given by
\begin{equation}
  \widetilde{\Omega}_i(A^\mu , \pi_\mu ; \Phi^j)
         =  \Omega_i + \sum_{n=1}^{\infty} \widetilde{\Omega}_i^{(n)}=0;
                       ~~~~~~\Omega_i^{(n)} \sim (\Phi^j)^n
\end{equation}
satisfying the boundary conditions,
$\widetilde{\Omega}_i( A^\mu , \pi_\mu;  0) = \Omega_i$.
Note that the modified constraints $\widetilde{\Omega}_{i}$
become strongly zero by introducing the auxiliary fields $\Phi^i$,
i.e., enlarging the phase space,
while the original constraints $\Omega_{i}$ are weakly zero.
As will be shown later, essentially due to this property, the result of the
Dirac formalism can be easily read off from the BFT formalism.
The first order correction terms in the infinite series $^{7}$ are
simply given by
\begin{equation}
  \widetilde{\Omega}_i^{(1)}(x) = \int d^3 y X_{ij}(x,y)\Phi^j(y),
\end{equation}
and the first class  constraint algebra (8) of $\widetilde{\Omega}_i$
requires
the following relation
\begin{equation}
   \triangle_{ij}(x,y) +
   \int d^3 w~ d^3 z~
        X_{ik}(x,w) \omega^{kl}(w,z) X_{jl}(y,z)
         = 0.
\end{equation}
However, as was emphasized in Refs. 13, 18, and 19,
there is a natural arbitrariness
in choosing the matrices $\omega^{ij}$ and $X_{ij}$ from Eqs. (7) and (10),
which corresponds to canonical transformation
in the extended phase space. $^{6,7}$
Here we note that Eq. (11) can not be
considered as the matrix multiplication exactly unless $X_{jl}(y,z)$ is
the symmetric matrix, i.e., $X_{jl}(y,z)=X_{lj}(z,y)$ because
of the form of the last two
product of the matrices $\int d^3z \omega^{kl}(\omega, z) X_{jl}(y,z)$ in
the right hand side of Eq. (11). Thus, using this arbitrariness
we can take the
simple solutions without any loss of generality, which are compatible with
Eqs. (7) and (11) as
\begin{eqnarray}
  \omega^{ij}(x,y)
         &=&  \epsilon^{ij} \delta^3({\bf x}-{\bf y}), \nonumber  \\
  X_{ij}(x,y)
         &=&  m \delta_{ij} \delta^3({\bf x}-{\bf y}),
\end{eqnarray}
i.e., antisymmetric $\omega^{ij}(x,y)$ and symmetric $X_{ij}(x,y)$ such that
Eq. (11) is the form of the matrix multiplication exactly $^{10, 13, 17-19}$
\begin{eqnarray*}
\Delta_{ij}(x,y) + \int d^3 \omega d^3 z X_{ik}(x, \omega) \omega^{kl}
(\omega, z) X_{lj}(z,y) =0. \nonumber
\end{eqnarray*}
Note that $X_{ij}(x,y)$ needs not be generally symmetric, while
$\omega^{ij}(x,y)$ is always antisymmetric by definition of Eq. (7).
However, the symmetricity of $X_{ij}(x,y)$ is, by experience, a
powerful property for
the solvability of (9) with finite iteration $^{13, 17-19}$ or with
infinite {\it regular } iterations. $^{24}$

In our model with this proper choice,
the modified constraints up to the first order iteration term
\begin{eqnarray}
\widetilde{ \Omega}_{i}&=&\Omega_{i}+\Omega^{(1)}_{i}  \nonumber \\
&=&\Omega_{i} +m \Phi^{i}
\end{eqnarray}
strongly form a first class  constraint algebra as follows
\begin{equation}
  \{ \Omega_{i}(x)+ \widetilde{\Omega}^{(1)}_{i}(x),~
           \Omega_{j}(y)+ \widetilde{\Omega}^{(1)}_{j}(y) \} = 0.
\end{equation}
Then, the higher order iteration terms
\begin{eqnarray}
\widetilde{\Omega}^{(n+1)}_{i} =
                  - \frac{1}{n+2} \Phi^{l}
                  \omega_{lk} X^{kj} B_{ji}^{(n)}~~~~~~~(n \geq 1)
\end{eqnarray}
with
\begin{eqnarray}
B^{(n)}_{ji} \equiv \sum^{n}_{m=0}
          \{ \widetilde{\Omega}^{(n-m)}_{j},
                \widetilde{\Omega}^{(m)}_{i} \}_{(A, \pi )}
        + \sum^{n-2}_{m=0}
              \{ \widetilde{\Omega}^{(n-m)}_{j},
                     \widetilde{\Omega}^{(m+2)}_{i} \}_{(\Phi)}
\end{eqnarray}
are found to be vanishing without explicit calculation.
Here, $\omega_{lk}$ and $X^{kj}$ are the inverse of
$\omega^{lk}$ and $X_{kj}$,
and the Poisson brackets including the subscripts are defined by
\begin{eqnarray}
\{A,B \}_{(q,p)} &\equiv& \frac{\partial A}{\partial q}
                  \frac{\partial B}{\partial p}
                - \frac{\partial A}{\partial p}
                  \frac{\partial B}{\partial q},  \nonumber \\
\{A,B \}_{(\Phi)} &\equiv& \sum_{i,j}
               \left[  \frac{\partial A}{\partial \Phi^{i}}
                       \frac{\partial B}{\partial \Phi^{j}}
                     - \frac{\partial A}{\partial \Phi^{j}}
                       \frac{\partial B}{\partial \Phi^{i}} \right],
\end{eqnarray}
where $(q,p)$ and $(\Phi^{i}, \Phi^{j} )$ are the conjugate pairs,
respectively.

\vspace{1cm}

\begin{center}
\subsection{\bf Physical Variables, First Class Hamiltonian, and Dirac
Brackets }
\end{center}

Now, corresponding to the original variables $A^{\mu}$ and $\pi_{\mu}$,
the ${\it physical }$ variables within the Abelian conversion in the extended
phase space,
$\widetilde{A}^{\mu}$ and $\widetilde{\pi}_{\mu}$
, which are strongly involutive, i.e.,
\begin{eqnarray}
\{ \widetilde{\Omega}_{i}, \widetilde{A}^{\mu} \} =0,~~~
\{ \widetilde{\Omega}_{i}, \widetilde{\pi}^{\mu} \} =0,
\end{eqnarray}
can be generally found as
\begin{eqnarray}
\widetilde{A}^{\mu}(A^{\nu}, \pi_{\nu}; \Phi^{j} ) &=&
          A^{\mu} + \sum^{\infty}_{n=1} \widetilde{A}^{\mu (n)},
        ~~~~~~~ \widetilde{A}^{\mu (n)} \sim (\Phi^{j})^{n};   \nonumber \\
\widetilde{\pi}_{\mu}(A^{\nu}, \pi_{\nu}; \Phi^{j} ) &=&
            \pi_{\mu} + \sum^{\infty}_{n=1} \widetilde{\pi}_{\mu}^{ (n)},
            ~~~~~~~ \widetilde{\pi}_{\mu}^{(n)} \sim (\Phi^{j})^{n}
\end{eqnarray}
satisfying the boundary conditions,
$\widetilde{A}^{\mu}(A^{\nu}, \pi_{\nu}; 0 ) = A^{\mu}$ and
$\widetilde{\pi}_{\mu}(A^{\nu}, \pi_{\nu}; 0 ) = \pi_{\mu}$.
Here, the first order iteration terms are given by
\begin{eqnarray}
\widetilde{A}^{\mu (1)}&=&
                   - \Phi^{j}\omega_{jk} X^{kl}
                   \{ \Omega_{l}, A^{\mu} \}_{(A, \pi)}  \nonumber  \\
              &=& ( \frac{1}{m} \Phi^{2}, m \partial_{i} \Phi^{1} ),
                              \nonumber  \\
\widetilde{\pi}_{\mu}^{(1)} &=& - \Phi^{j}\omega_{jk} X^{kl}
                   \{ \Omega_{l}, \pi_{\mu} \}_{(A, \pi)}  \nonumber \\
              &=&  ( m \Phi^{1}, 0 ).
\end{eqnarray}
Furthermore, since the modified variables up to the first iterations,
$A^{\mu}+\widetilde{A^{\mu}}^{(1)}$
and $\pi_{\nu} +\widetilde{\pi}_{\nu}^{(1)}$ are found to involutive,
i.e., to satisfy Eq. (18), the higher order iteration terms
\begin{eqnarray}
\widetilde{A}^{\mu (n+1)} &=&
                              -\frac{1}{n+1}
                              \Phi^{j}\omega_{jk} X^{kl} (G^\mu)^{(n)}_l,
                              \nonumber \\
\widetilde{\pi}_{\mu}^{(n+1)} &=& -\frac{1}{n+1} \Phi^{j}
                                  \omega_{jk} X^{kl} (H_{\mu})^{(n)}_{l}
\end{eqnarray}
with
\begin{eqnarray}
(G^\mu)^{(n)}_l &=& \sum^{n}_{m=0}
                \{ \Omega_{i}^{(n-m)}, \widetilde{A}^{\mu (m)}\}_{(A, \pi )}
                +   \sum^{n-2}_{m=0}
                \{ \Omega_{i}^{(n-m)}, \widetilde{A}^{\mu (m+2)}\}_{(\Phi)}
           +  \{ \Omega_{i}^{(n+1)}, \widetilde{A}^{\mu (1)} \}_{(\Phi)},
                     \nonumber  \\
{(H_{\mu})}^{(n)}_{l} &=& \sum^{n}_{m=0}
        \{ \Omega_{i}^{(n-m)}, \widetilde{\pi}_{\mu}^{(m)} \}_{(A, \pi )}
     + \sum^{n-2}_{m=0}
           \{ \Omega_{i}^{(n-m)}, \widetilde{\pi}_{\mu}^{(m+2)} \}_{(\Phi )}
     + \{ \Omega_{i}^{(n+1)}, \widetilde{\pi}_{\mu}^{(1)} \}_{(\Phi )}
                     \nonumber \\
\end{eqnarray}
are also found to be automatically vanishing.
Hence, the physical variables in the extended phase space
are finally found to be
\begin{eqnarray}
\widetilde{A}^{\mu} &=& A^{\mu} + \widetilde{A}^{\mu (1)}  \nonumber \\
                    &=& (A^{0} + \frac{1}{m} \Phi^{2},
                        A^{i} + \frac{1}{m} \partial^{i} \Phi^{1} ),  
\nonumber \\
\widetilde{\pi}_{\mu} &=& \pi_{\mu} + \widetilde{\pi}_{\mu}^{(1)}
                                                            \nonumber \\
                      &=& (\pi_{0}+m \Phi^{1}, \pi_{i} )  \nonumber \\
                     &=&(\widetilde{\Omega}_{1}, \pi_{i}).
\end{eqnarray}
Similar to the physical phase space variables $\widetilde{A}^{\mu}$ and
$\widetilde{\pi}_{\mu}$, all
other physical quantities, which correspond to the functions of
$A^{\mu}$ and $\pi_{\mu}$,
can be also found in principle by considering the solutions like
as Eq. (19). $^{7, 13, 17-19}$ However, it is expected that this procedure
of finding
the physical quantities may not be simple
depending on the complexity of the functions.

In this paper, we consider a new approach using the property $^{7,25}$
\begin{eqnarray}
\widetilde{K}(A^{\mu}, \pi_{\mu}; \Phi^{i} )= K(\widetilde{A}^{\mu},
\widetilde{\pi}_{\mu} )
\end{eqnarray}
for the arbitrary function or functional $K$ defined on the original phase
space variables unless $K$ has the time derivatives:
the following relation
\begin{eqnarray}
\{ K(\widetilde{A}^{\mu}, \widetilde{\pi}_{\mu} ),\widetilde{\Omega}_{i} \}=0
\end{eqnarray}
is satisfied for
any function $K$ not having the time derivatives
because $\widetilde{A}^{\mu}$ and $\widetilde{\pi}_{\mu}$ and their spatial
derivatives already
commute with $\widetilde{\Omega}_{i}$ at equal times by definition.
On the other hand, since the solution $K$ of Eq. (24) is
unique up to the power of the first class constraints
$\widetilde{\Omega}_{i}$, $^{17,19}$
$K(\widetilde{A}^{\mu}, \widetilde{\pi}_{\mu})$ can be identified
with $\widetilde{K}(A^{\mu}, \pi_{\mu}; \Phi)$ modulus the power of the first
class constraints $\tilde{\Omega}_{i}$.
However, note that this property is not satisfied when
the time derivatives exist.

Using this elegant property
we can directly obtain the desired first class Hamiltonian
$\widetilde{H_{T}}$
corresponding to the total Hamiltonian $H_{T}$ of Eq. (3)
as follows
\begin{eqnarray}
\widetilde{H_{T}}(A^{\mu}, \pi_{\nu}; \Phi^{i} )
    &=&    H_{T}(\widetilde{A}^{\mu}, \widetilde{\pi}_{\nu})\nonumber \\
    &=&    H_{T}(A^{\mu}, \pi_{\nu})+
              \int d^{3}x \left[
                  \frac{1}{2}( \partial_{i} \Phi^{1} )^{2}
                 +\frac{1}{2} (\Phi^{2})^{2}
                 + \frac{1}{m}\partial_{i} \partial_{i} \Phi^{1} 
\widetilde{\Omega_{1}}
                 - \frac{1}{m} \Phi^{2}
                     \widetilde{\Omega}_{2}
                                \right]. \nonumber \\
\end{eqnarray}
On the other hand, since the first class Hamiltonian
$\widetilde{H_{c}}$ corresponding
to the canonical Hamiltonian $H_{c}$ of Eq. (4) can be similarly obtained as
follows
\begin{eqnarray}
\widetilde{H_{c}}(A^{\mu}, \pi_{\nu}; \Phi^{i} )
    &=&    H_{c}(\widetilde{A}^{\mu}, \widetilde{\pi}_{\nu})\nonumber \\
    &=&    H_{c}(A^{\mu}, \pi_{\nu})+
              \int d^{3}x \left[
                 \frac{1}{2}( \partial_{i} \Phi^{1} )^{2}
                + \frac{1}{2} (\Phi^{2})^{2}
                + m \Phi^{1} \partial^{i} A_{i}
                 - \frac{1}{m} \Phi^{2}
                     \widetilde{\Omega}_{2}
                                \right], \nonumber \\
\end{eqnarray}
Eq. (26) can be re-expressed as follows
\begin{eqnarray}
\widetilde{H_{T}}(A^{\mu}, \pi_{\nu}; \Phi^{i} ) =
  \widetilde{H_{c}}(A^{\mu}, \pi_{\nu}; \Phi^{i} ) -\int d^{3}x(
\partial_{i} \widetilde{A^{i}}) \widetilde{\Omega}_{1}
\end{eqnarray}
as it should be according to Eq. (3).

This is the same result as the usual approach (See the Appendix {\bf A})
and, by construction, both $\widetilde{H_{T}}$ and $\widetilde{H_{c}}$ are
automatically strongly involutive,
\begin{eqnarray}
\{ \widetilde{\Omega}_{i}, \widetilde{H_{T}} \} &=&0, \\ \nonumber
\{ \widetilde{\Omega}_{i}, \widetilde{H_{c}} \} &=&0.
\end{eqnarray}
Note that all our constraints have already this property, i.e.,
$\widetilde{\Omega}_{i}(A^{\mu}, \pi_{\mu}; \Phi)
=\Omega_{i}(\widetilde{A}^{\mu}, \widetilde{\pi}_{\mu})$.
In this way, the second class constraints system
$\Omega_{i}(A_{\mu}, \pi^{\mu} ) \approx 0$ is converted into
the first class constraints one
$\widetilde{\Omega}_{i}(A_{\mu}, \pi^{\mu}; \Phi ) =0$
with the boundary conditions
$\widetilde{\Omega}_{i} |_{\Phi=0}=\Omega_{i}$.

On the other hand, in the Dirac formalism $^{1}$
one can make the second class constraint system
$\Omega_{i} \approx 0$ into the first class constraint one
$\Omega_{i}(A_{\mu}, \pi^{\mu} )=0$ only by deforming the phase space
$(A_{\mu}, \pi^{\mu})$ without introducing any new fields.
Hence, it seems that
these two formalisms are drastically different ones.
However, remarkably the
Dirac formalism can be easily read off from the usual BFT-formalism $^{7}$
by noting that the Poisson bracket in the extended phase space with
$\Phi \rightarrow 0$ limit becomes
\begin{eqnarray}
\{ \widetilde{A}, \widetilde{B} \} |_{\Phi=0}
              &=&   \{A, B \}
                  - \{A, \Omega_{k} \} \Delta^{kk'} \{\Omega_{k'}, B
\}  \nonumber \\
              &=&   \{A, B \}_{D}
\end{eqnarray}
where $\Delta^{kk'}=-X^{lk} \omega_{ll'} X^{l'k'}$ is the inverse of
$\Delta_{kk'}$ in Eq. (6). About this remarkable relation, we note that this
is essentially due to the Abelian conversion method of the original second
class constraint: In this case the Poisson brackets between the
constraints and
the other things in the extended phase space are already strongly zero
\begin{eqnarray}
\{ \widetilde{\Omega}_{i}, \widetilde{A} \} &=&0 , \nonumber \\
\{ \widetilde{\Omega}_{i}, \widetilde{\Omega}_{j} \} &=&0,
\end{eqnarray}
which resembles the property of the Dirac bracket in the non-extended phase
space
\begin{eqnarray}
\{\Omega_{i}, A \}_{D} &=&0, \nonumber \\
\{\Omega_{i}, \Omega_{j} \}_{D} &=&0,
\end{eqnarray}
such that
\begin{eqnarray}
\{ \widetilde{\Omega}_{i}, \widetilde{A} \} |_{\Phi=0}
          &\equiv& \{ \Omega_{i}, A \}^{*} =0, \nonumber \\
\{ \widetilde{\Omega}_{i}, \widetilde{\Omega}_{j} \} |_{\Phi=0}
           &\equiv& \{ \Omega_{i}, \Omega_{j} \}^{*} =0
\end{eqnarray}
are satisfied for some bracket in the non-extended phase space 
$\{~~,~~\}^{*}$.
However, due to the uniqueness of the Dirac bracket $^{26}$ it is natural
to expect the previous result (29) is satisfied, i.e.,
\begin{eqnarray}
\{~~,~~\}^{*} =\{~~,~~\}_{D}
\end{eqnarray}
without explicit manipulation. Moreover we add that, due to similar reason,
some non-Abelian generalization of the Abelian conversion as
\begin{eqnarray}
\{\widetilde{\Omega}_{i}, \widetilde{A} \} &=&\alpha_{ij} \Phi^{j} 
+\alpha_{ijk} \Phi^{j} \Phi^{k} + \cdots ,\nonumber \\
\{\widetilde{\Omega}_{i}, \widetilde{\Omega}_{j} \} &=&\beta_{ijk} \Phi^{k} +
    \beta_{ijkl} \Phi^{k} \Phi^{l} + \cdots
\end{eqnarray}
also gives the same result (30) with the functions $\alpha_{ij},
\alpha_{ijk},
 \beta_{ijk}, etc, \cdots $ of the original phase variables $A_{\mu},
~\pi^{\nu}$.
As an specific example,
let us consider the brackets between the phase space variables in Eq. (23).
The results are as follows
\begin{eqnarray}
&&\{\widetilde{A}^{0}(x), \widetilde{A}^{j}(y) \} |_{\Phi=0}
         = \{ \widetilde{A}^0 (x), \widetilde{A}^j (y) \}
         = \frac{1}{m^{2}} \partial_{x}^{j} \delta({\bf x}-{\bf y}),
                                                           \nonumber \\
&&\{\widetilde{A}^{0}(x), \widetilde{A}^{0}(y) \} |_{\Phi=0}
         = \{ \widetilde{A}^0 (x), \widetilde{A}^0 (y) \}=0,
                                                           \nonumber  \\
&&\{\widetilde{A}^{j}(x), \widetilde{A}^{k}(y) \} |_{\Phi=0}
         = \{ \widetilde{A}^j (x), \widetilde{A}^k (y) \}=0,
                                                           \nonumber \\
&&\{\widetilde{\Pi}_{\mu}(x), \widetilde{\Pi}_{\nu}(y) \} |_{\Phi=0}
         = \{ \widetilde{\Pi}_\mu (x), \widetilde{\Pi}_\nu (y) \}=0,
                                                           \nonumber \\
&&\{\widetilde{A}^{i}(x), \widetilde{\Pi}_{j}(y) \} |_{\Phi=0}
         = \{ \widetilde{A}^i (x), \widetilde{\Pi}_j (y) \}
         = \delta_{ij} \delta (\bf{x}-\bf{y}),
                                                           \nonumber \\
&&\{\widetilde{A}^{0}(x), \widetilde{\Pi}_{\nu}(y) \} |_{\Phi=0}
         = \{ \widetilde{A}^0 (x), \widetilde{\Pi}_\nu (y) \}=0,
                                                           \nonumber \\
&&\{\widetilde{A}^{i}(x), \widetilde{\Pi}_{0}(y) \} |_{\Phi=0}
         = \{ \widetilde{A}^i (x), \widetilde{\Pi}_0 (y) \}=0
\end{eqnarray}
due to the linear correction (i.e., only the first order correction) of the
modified fields $\widetilde{A}^{\mu}$ and $\widetilde{\pi}_{\mu}$,
which are the same as the usual Dirac brackets. $^{19}$
Hence, in the case of the phase space variables of the model the well-known
Dirac brackets of the fields are
exactly the Poisson brackets of the corresponding modified fields in the
extended phase space contrast to the general formula (30).
Note that the FJ symplectic formalism, $^{22}$
which is the improved version of the Dirac method,
also gives the same result (See Appendix ${\bf B}$ for this matter).

Now, since in the Hamiltonian formalism the first class constraint system
without the CS term $^{13,18}$ indicates the presence of a local symmetry,
this  completes  the  operatorial conversion of the original
second class system with the Hamiltonian $H_T$
and the constraints $\Omega_i$
into first class one with the Hamiltonian $\widetilde{ H_T}$
and the constraints $\widetilde{\Omega}_i$.
From Eqs. (13) and (28),
one can easily see that the original second class constraint system
is converted into the effectively first class one if one introduces
two fields,
which are conjugated with each other in the extended phase space.
Note that for the Proca case
the origin of the second class constraint
is due to the explicit gauge symmetry breaking term in the action (1).

\vspace{1cm}
\begin{center}
\subsection{Corresponding First Class Lagrangian}
\end{center}

Next, we consider the partition function of the model
in order to present the Lagrangian  corresponding to $\widetilde{H_{T}}$ in
the canonical Hamiltonian formalism. However, the result is the same with
$\widetilde{H_{c}}$.
As a result, we will unravel the correspondence of
the Hamiltonian approach with
the well-known St\"uckelberg's formalism.
First, let us identify the new variables
$\Phi^i$ as a canonically conjugate pair
($\rho$, $\pi_\rho$) in the Hamiltonian formalism,
\begin{equation}
(\Phi^i) \to  (m \rho, \frac{1}{m}\pi_\rho),
\end{equation}
satisfying Eqs. (7) and (12).
Then, the starting phase space partition function is given by the Faddeev
formula $^{3,27}$ as follows
\begin{equation}
Z=  \int  {\cal D} A^\mu
          {\cal D} \pi_\mu
          {\cal D} \rho
          {\cal D} \pi_\rho
               \prod_{i,j = 1}^{2} \delta(\widetilde{\Omega}_i)
                           \delta(\Gamma_j)
               \mbox{det} \mid \{ \widetilde{\Omega}_i, \Gamma_j \} \mid
                e^{iS},
\end{equation}
where
\begin{equation}
S  =  \int d^4x \left(
         \pi_\mu {\dot A}^\mu + \pi_\rho {\dot \rho} 
 - \widetilde {\cal H}_{T}
                \right),
\end{equation}
with the Hamiltonian density $\widetilde {\cal H}_{T}$
corresponding to the Hamiltonian
$\widetilde H_{T}$ of Eq. (25), which is now expressed in terms
of $(\rho, \pi_\rho)$ instead of $\Phi^i$.
Note that the gauge fixing conditions $\Gamma_i$ are chosen
so that the determinant occurring in
the functional measure is nonvanishing.
Moreover, $\Gamma_i$ may be assumed to be independent of the momenta
so that these are considered as
the Faddeev-Popov type gauge conditions. $^{10,13,17,27,28}$

Before performing the momentum integrations to obtain the
partition function in the configuration space,
it seems appropriate to comment on the involutive Hamiltonian.
If we directly use the above Hamiltonian following
the previous analysis done
by Banerjee {\it {et al.}}, $^{19}$
we will finally obtain the non-local action corresponding
to this Hamiltonian
due to the existence of $(\partial^i \pi_i)^2$--term in the action
when we carry out the functional integration over $\pi_{\rho}$ later.
Furthermore, if we use the above Hamiltonian,
we can not also naturally generate
the first class Gauss' law constraint $\widetilde{\Omega}_2$ from
the time evolution of the primary constraint
$\widetilde{\Omega}_1$, i.e., $\{\widetilde{\Omega}_{1}, 
\widetilde{H_{T}} \}=0$.
Therefore, in order to avoid these unwanted situations,
we use the equivalent first class Hamiltonian
without any loss of generality,
which differs from the involutive Hamiltonian (26)
by adding a term proportional to the first class constraint
$\widetilde{\Omega}_2$ as follows
\begin{equation}
\widetilde{H_{T}}' = \widetilde {H_{T}} + \int d^{3}
x\frac{\pi_\rho}{m^2} \widetilde{\Omega}_2.
\end{equation}
Then, we have the desired first constraint system such that
\begin{eqnarray}
\{ \widetilde{\Omega}_1, \widetilde{H_{T}}'\}
                         &=& \widetilde{\Omega}_2, \nonumber \\
\{ \widetilde{\Omega}_2, \widetilde{H_{T}}'\}
                         &=& 0.
\end{eqnarray}
Note that when we operate this modified Hamiltonian on physical states,
the difference is trivial because such states are
annihilated by the first class constraints.
Similarly, the equations of motion for gauge invariant variables
will also be unaffected by this difference
since $\widetilde{\Omega}_2$ can be regarded as
the generator of the gauge transformations.

Now, we consider the following effective phase space partition function
\begin{eqnarray}
Z &=& \int {\cal D} \pi_{\mu} {\cal D} A^{\mu}
           {\cal D} \pi_{\rho}
           {\cal D} \rho
           \prod^2_{i,j = 1}
            \delta(\widetilde{\Omega}_i) \delta(\Gamma_j)
            \mbox{det} \mid \{ \widetilde{\Omega}_i,  \Gamma_j \} \mid
            e^{iS'}, \nonumber \\
S' &=& \int  d^4 x ~( \pi_{\mu} {\dot A^{\mu}} +
            \pi_{\rho} \dot{\rho}
            -  \widetilde{\cal H}_{T}').
\end{eqnarray}
The $\pi_0$ integral is performed trivially by exploiting the delta function
$\delta(\widetilde{\Omega}_1) = \delta(\pi_0 + m^2 \rho)$ in Eq. (42).
On the other hand, the other delta function
$\delta(\widetilde{\Omega}_2) = \delta
(\partial^i \pi_i + m^2 A^0 + \pi_\rho)$ can be expressed
by its Fourier transform with Fourier variable $\xi$ as follows
\begin{equation}
\delta(\widetilde{\Omega}_2) = \int {\cal D}\xi
                          e^{-i \int d^4x ~\xi \widetilde{\Omega}_2}.
\end{equation}
Making a change of variable $A^0 \to A^0 + \xi$, we obtain the action
\begin{eqnarray}
S &=& \int d^4x~ [ \pi_i {\dot A}^i - m^2 \rho ( \dot{A^0} + \dot{\xi} )
                    + \pi_{\rho}\dot{\rho}
                - \frac{1}{2} \pi_i^2 - \frac{1}{4} F_{ij} F^{ij}
                + \frac{1}{2} m^2 (A^0)^2
                + \frac{1}{2} m^2 A_i A^i \nonumber\\
   &+& A^0 \partial^i \pi_i
                 - m^2 \partial_i A^i \rho
                - \frac{1}{2m^2} \pi_\rho^2
                + \frac{1}{2} m^2 \partial_i \rho \partial^i \rho
                - \xi \pi_\rho - \frac{1}{2} m^2 \xi^2 ],
\end{eqnarray}
where the corresponding measure is given by
\begin{equation}
[{\cal D} \mu] = {\cal D} \xi
                 {\cal D} \pi_i
                 {\cal D} A^\mu
                {\cal D} \rho
                {\cal D} \pi_\rho
                \prod_j
               \{ \delta[ \Gamma_j (A^0+\xi, A^i, \pi_i, \rho) ] \}
               \mbox{det} \mid \{\widetilde{\Omega}_i, \Gamma_j \} \mid.
\end{equation}
Performing the Gaussian integral over $\pi_i$, this yields
the intermediate action as follows
\begin{eqnarray}
S_u &=&  \int d^4x~ [ - \frac{1}{4} F_{\mu\nu} F^{\mu\nu}
                    + \frac{1}{2} m^2 A_\mu A^\mu          \nonumber\\
                  &+& \pi_\rho ( \dot{\rho} - \xi - \frac{1}{2m^2} \pi_\rho)
                   - m^2 \rho {(\dot{A^0} + \dot{\xi} )}
                   - m^2 \partial_i A^i \rho
                   + \frac{1}{2} m^2 \partial_i\rho \partial^i\rho
                   - \frac{1}{2} m^2 \xi^2].
\end{eqnarray}
To realize the St\"uckelberg term through the BFT analysis,
we choose the Faddeev-Popov-like gauges
$^{10,13,17,27,28}$,
which do not involve the momenta.
After the Gaussian integration over $\pi_\rho$,
we finally obtain the well-known action up to the total divergence
by identifying the extra field $\rho$ with the St\"uckelberg scalar
as follows
\begin{equation}
S = \int d^4x  [ - \frac{1}{4}  F_{\mu\nu} F^{\mu\nu}
                 + \frac{1}{2} m^2 (A_\mu + \partial_\mu \rho)^2],
\end{equation}
which is invariant under the gauge transformations as
$\delta A_\mu = \partial_\mu \Lambda$ and $\delta \rho = - \Lambda$.
As expected, the St\"uckelberg scalar $\rho$ is introduced in the mass term.

It seems to appropriate to comment on the original unitary gauge fixing.
If we choose the gauge as follows
\begin{equation}
\Gamma_{i} = (\rho, ~ \pi_\rho ),
\end{equation}
we get
\begin{equation}
\{ \widetilde{\Omega}_i({\bf x}), \Gamma_j ({\bf y}) \}
=\epsilon_{ij}\delta^3({\bf x}-{\bf y}).
\end{equation}
Then, integrating over the variables $\rho$ and $\pi_\rho$
we reproduce the original partition function as follows
\begin{equation}
{\cal Z} = \int {\cal D} A_\mu e^{i\int~ d^4x ~
          [-\frac{1}{4} F_{\mu\nu} F^{\mu\nu}
          - \frac{1}{2} m^2 A_\mu A^\mu]}.
\end{equation}
The physical meaning of this result is that the original action
can be regarded as a gauge-fixed version of the first constraint
system (18) and (40).
Note that this gauge fixing is consistent because when we take the
gauge fixing condition $\rho \approx 0$,
the condition $\pi_\rho \approx 0$ is
naturally generated from the time evolution of $\rho$, i.e.,
$\dot{\rho} = \{\rho , H_u \}=-\frac{1}{m^2}\pi_\rho \approx 0$,
where the Hamiltonian $H_u$ corresponds to the intermediate action $S_u$.

\vspace{1cm}
\begin{center}
\section {\bf BFV-BRST Gauge Fixing}
\end{center}

\begin{center}
\subsection {\bf Basic Structure}
\end{center}

In this subsection, we briefly recapitulate
the BFV formalism $^{2,3}$
which is applicable for general theories
with the first-class constraints.
For simplicity, this formalism is restricted to a finite
number of phase space variables. This makes the discussion simpler and
conclusions more apparent.

First of all, consider a phase space of the bosonic canonical
variables $q^{i},~p_{i}$
($i$ = 1, 2, $\cdots$, n) in terms of which canonical Hamiltonian
$H_{c}(q^{i},p_{i})$ and constraints $\Omega_{a}(q^{i},p_{i}) \approx 0$
($a$ = 1, 2, $\cdots$, m), which being bosonic also, are given. We
assume that the constraints
satisfy the following constraint algebra $^{2,3,9}$
\begin{eqnarray}
\{\Omega_{a}, \Omega_{b} \}
                &=& U^{c}_{a b} \Omega_{c} ~, \nonumber \\
\{H_{c}, \Omega_{a} \}
                &=& V^{b}_{a} \Omega_{b} ~,
\end{eqnarray}
where the structure coefficients $U^{c}_{a b}$ and $V^{b}_{a}$ are
functions of the canonical variables. We also assume that the constraints
are irreducible, which means that there locally exists an invertible
change of
the variables such that $\Omega_{a}$ can be identified
with the $m$-unphysical momenta.

In order to single out the physical variables, we can introduce the
additional bosonic conditions $\Gamma^{a}(q^{i},p_{i}) \approx 0$ with
$\det |\{ \Gamma^{a}, \Omega_{b} \}| \neq 0$ at least in the vicinity
of the constraint surface $\Gamma^{a} \approx 0$ and $\Omega_{a} \approx 0$.
Then, $\Gamma^a$ play the roles of gauge-fixing functions. That is to say,
from the condition of time stability of the constraints, a family of phase
space trajectories is possible. By selecting one of these
trajectories through the conditions of $\Gamma^{a} \approx 0$, we can get
the 2($n-m$) dimensional physical sub-phase space
denoted by $q^{*},~p^{*}$. And then, $\Gamma^{a} (q^{i},p_{i})$ can be
identified with the $m$-unphysical coordinates.

The described dynamical system with the partition function
\begin{equation}
{\cal Z} = \int \!  [ d q^{i} d p_{i} ]~ \delta ( \Omega_{a} )
             \delta ( \Gamma^{b} ) \det |\{ \Gamma^{b}, \Omega_{a} \}|~
              e^{ i\!\!\int \!\! d x(p \dot{q} - H_{c})}
\end{equation}
is then completely equivalent to the effective quantum theory with the
following partition function
\begin{equation}
{\cal Z}_{eq.}
    = \int \!  [d q^{*} d p^{*} ]~  e^{ i\!\!\int \!\! d x
      [ p^{*} \dot{q}^{*} - H_{phys} (q^{*},p^{*}) ]}
\end{equation}
which only depending on the canonical variables $q^*, p^*$ of the physical
sub-phase space,

And the constraints $\Omega_{a} \approx 0$ and $\Gamma^{a} \approx 0$
together with the Hamilton equations may be obtained from a action
\begin{equation}
S=\int \! d t ~(p_{i} \dot{q}^{i}-H_{c}-\lambda^{a} \Omega_{a} +\pi_{a}
   \Gamma^{a} )~,
\end{equation}
where $\lambda^{a}$ and $\pi_{a}$ are the bosonic Lagrange
multiplier fields canonically conjugated to each other, obeying the
Poisson bracket relations
\begin{equation} \{\lambda^{a}, \pi_{a}\}=\delta^{a}_{b}~.
\end{equation}
Note that the gauge-fixing
conditions contain $\lambda^{a}$ in the following general form
\begin{equation}
   \Gamma^{a}=\dot{\lambda}^{a}+\chi^{a}(q^{i},p_{i},\lambda^{a}),
\end{equation}
where $\chi^{a}$ are arbitrary functions. And we can see
that the Lagrange multiplier $\lambda^{a}$ become dynamically active, and
$\pi_{a}$ serve as their conjugate momenta. This consideration naturally
leads to the canonical formalism in an extended phase space.

In order to make the equivalence to the initial theory
with the constraints in the reduced phase space,
we may introduce two sets of canonically conjugate, fermionic ghost
coordinates and momenta ${\cal C}^{a},~\overline{\cal P}_{a}$ and
${\cal P}^{a},~\overline{\cal C}_{a}$ such that
\begin{equation}
\{ {\cal C}^{a}, \overline{\cal P}_{b} \}~=~\{ {\cal P}^{a},
\overline{\cal C}_{b} \}~=~\delta^{a}_{b}~
\end{equation}
with the super-Poisson bracket
\begin{eqnarray*}
\{ A, B \} =\left. \frac{\delta A}{\delta q} \right|_{r}
            \left. \frac{\delta B}{\delta p} \right|_{l}
      -(-1)^{\eta_{A} \eta_{B}}
            \left. \frac{\delta B}{\delta q} \right|_{r}
            \left. \frac{\delta A}{\delta p} \right|_{l},
\end{eqnarray*}
where $\eta_{A}$ denotes the number of fermions called ghost number in $A$,
and subscript ``$r$'' and ``$l$'' right and left derivatives.

The quantum theory is now defined by the extended phase space functional
integral
\begin{equation}
{\cal Z}_{\Psi}~=~ \int [d q^{i} d p_{i}][d \lambda^{a} d \pi_{a}]
         [ d {\cal C}^{a} d {\overline{\cal P}}_{a} ]
         [ d {\cal P}^{a} d {\overline{\cal C}}_{a} ]~
          e^{iS_{\Psi}}~,
\end{equation}
where the action is
\begin{equation}
S_{\Psi}=\int \! d t \{p_{i} \dot{q}^{i}+\pi_{a} \dot{\lambda}^{a}+
           {\overline{\cal P}}_{a} \dot{\cal C}^{a}+
           {\overline{\cal C}}_{a} \dot{\cal P}^{a} -
           H_{m}+\{Q,\Psi \} \}.
\end{equation}
Here, the BRST-charge $Q$ and the fermionic gauge-fixing function $\Psi$
are defined by
\begin{eqnarray}
Q &=& {\cal C}^{a}\Omega_{a}~-~\frac{1}{2} {\cal C}^{b} {\cal C}^{c}
    U^{a}_{c b} {\overline{\cal P}}_{a}~+~{\cal P}^{a} \pi_{a}~, \nonumber \\
\Psi &=& {\overline{\cal C}}_{a} \chi^{a}~+~ {\overline{\cal P}}_{a} 
\lambda^{a}~,
\end{eqnarray}
respectively.
$H_{m}$ is the BRST invariant Hamiltonian, called the minimal Hamiltonian,
\begin{equation}
H_{m}~=~H_{c}~+~{\cal C}^{a} V_{a}^{b} ~{\overline{\cal P}}_{b}.
\end{equation}

The measure in ${\cal Z}_{\Psi}$ is the Liouville measure on the covariant
phase space. Furthermore, if we choose the fermionic
gauge-fixing function $\Psi$ properly, $^{3,11}$ we can obtain
manifestly covariant
expression. And the equivalence of the dimensionality $2n+6m$ in the
extended phase space, including the canonical ghost variables, to the
original dimensionality $2n-2m$ in the reduced phase space
can be seen by identifying the ghost variables
with the
negative-dimensional canonical degree of freedom, which is suggested
by the Parisi-Sourlas' original work related with the superrotation
Osp(1,1$\mid$2) in the extended phase space. $^{29}$

In order to derive the BRST gauge-fixed covariant action
for the Abelian Proca theory,
according to the above BFV formalism in the extended phase space,
let us introduce the ghosts and anti-ghosts
together with auxiliary fields as follows
\begin{equation}
({\cal C}^i , {\overline{\cal P}}_i),~({\cal P}^i , {\overline{\cal C}}_i ),
~(N^i , B_i)~,
\end{equation}
where $i = 1,~2$.
The nilpotent BRST-charge $Q$,
the fermionic gauge-fixing function $\Psi$ and the minimal
Hamiltonian $H_m$ in our case are
\begin{eqnarray}
Q~&=&~\int\!dx~[~{\cal C}^i \widetilde{\Omega}_i ~+~
                  {\cal P}^i B_i ~], \nonumber \\
\Psi~&=&~\int\!dx~[~{\overline{\cal C}}_i \chi^i 
                    ~+~
                     {\overline{\cal P}}_i N^i ~], \nonumber \\
H_{m}~&=&~\widetilde{H_{T}}'~-~\int\!dx
          ~[\overline{\cal P}_{2} {\cal C}^{1}],
\end{eqnarray}
where $\chi^1 =A^0 ,~\chi^2 = \partial_i A^i + \frac{\textstyle \alpha}
{\textstyle 2}B_2$, and $\alpha$ is an arbitrary parameter.

The BRST-charge $Q$, the fermionic gauge-fixing function $\Psi$, and
the minimal Hamiltonian $H_m$ satisfy the following relations,
\begin{eqnarray}
\{Q, H_{m} \}&=& 0, \nonumber \\
Q^{2}~=~\{Q,Q\}&=& 0, \nonumber \\
\{~\{ \Psi,Q\}, Q\}&=&0~,
\end{eqnarray}
which being the conditions of physical subspace after the operator 
quantization
\begin{eqnarray}
\left[\hat{A}, \hat{B} \right] = i \hbar \{A, B \}
\end{eqnarray}
for the quantum operators $\hat{A}$ and $\hat{B}$ corresponding to the
classical functions $A$ and $B$ when there is no operator ordering problem.

The effective action is
\begin{equation}
S_{eff} = \int\!d^4\!x~[~\pi_0 \dot{A}^0 + \pi_i \dot{A}^i
                    + \pi_\rho \dot{\rho}
                    + B_2\dot{N}^2+ \overline{\cal P}_i\dot{\cal C}^i
                    + \overline{\cal C}_2 \dot{\cal P}^2~ ] - H_{total}
\end{equation}
where $ H_{total} = H_{m} +\{Q,\Psi \}$, and also $\int d^{4}x(B_{1} 
\dot{N}^{1} +
\overline{\cal C}_{1} \dot{\cal P}^1)=\{ Q, \int d^{4}x~ \overline{\cal C}
\dot{N^{1}} \}$ terms are suppressed by replacing $\chi^{1}$ with
$\chi^{1} +\dot{N^{1}}$ just like the cases in Refs. 8, 9.

\vspace{1cm}
\begin{center}
\subsection {\bf Local Effective Action}
\end{center}

Now, in order to derive the covariant effective action we first
perform the path integration over
the fields $B_1 ,~N^1 ,~\overline{\cal C}_1 , {\cal P}^1 ,
~\overline{\cal P}_1 , {\cal C}^1 ,A^0$, and $\pi_0$ by  using of
the Gaussian integration. Then, we obtain
\begin{eqnarray}
S_{eff}&=&\int\!d^{4}x~
      [~\pi_i \dot{A}^i
      ~+~ \pi_{\rho} \dot{\rho}
      ~+~B\dot{N}
      ~+~{\overline{\cal P}}\dot{\cal C}
      ~+~{\overline{\cal C}}\dot{\cal P} \nonumber \\
   &&~~~-~\frac{1}{2}(\pi_i)^2
       ~-~ \frac{1}{2m^2}(\pi_{\rho})^2
       ~-~ \frac{1}{4} F_{ij} F^{ij}
       ~-~\frac{1}{2} m^{2} (A^i)^2
                                              \nonumber \\
   &&~~
       ~-~ m^2 \rho \partial_i A^i
       ~+~ \frac{1}{2} m^2 \partial_i \rho \partial^i \rho
       ~-~N(\partial^i \pi_i + \pi_\rho)  \nonumber \\
   &&~~
       ~-~B(\partial_{i}A^i ~+~ \frac{1}{2}\alpha B)
       ~-~\partial_{i}{\overline{\cal C}}
             \partial^{i}{\cal C} ~-~ {\overline{\cal P}}{\cal P}~]
\end{eqnarray}
with $N^2 \equiv N,~B_2 \equiv B,~
{\overline{\cal C}}_2 \equiv{\overline{\cal C}},~
{\cal C}^2 \equiv{\cal C},~
{\overline{\cal P}}_2\equiv{\overline{\cal P}}$, and
 ${\cal P}^2 \equiv{\cal P}.$
Using the variations over $\pi^{i},~\pi_{\rho},~{\cal P}$ and
${\overline{\cal P}}$, we obtain the following relations
\begin{eqnarray}
\pi_{i} &=& \dot{A}^i ~-~ \partial_i N, \nonumber \\
\pi_{\rho} &=& m^2 (\dot{\rho} - N), \nonumber \\
{\overline{\cal P}} &=& -\dot{\overline{\cal C}},
~~{\cal P}~~=~~\dot{\cal C},
\end{eqnarray}
and identifying with $N=-A^0$, we get the usual local form of the covariant
effective action
\begin{equation}
S_{eff} = \int\!d^4x~[ -\frac{1}{4}F_{\mu \nu}F^{\mu \nu}
            + \frac{1}{2} m^{2}
               ( A_{\mu} + \partial_\mu \rho )^2
            + A^{\mu}\partial_{\mu}B - \frac{1}{2}\alpha (B)^2
             - \partial_{\mu}{\overline{\cal C}}\partial^{\mu}{\cal C}~],
\end{equation}
which is invariant under the standard BRST transformation
\begin{eqnarray}
\delta_{B}A_{\mu} &=& -\lambda\partial_{\mu}{\cal C},
            ~~~~~~\delta_{B} \rho = \lambda {\cal C},  \nonumber \\
 \delta_{B}{\cal C}  &=& 0,
          ~~~~~~\delta_{B}{\overline{\cal C}} = - \lambda B,
          ~~~~~~\delta_{B}B = 0,
\end{eqnarray}
which is local and covariant one. This completes the procedure of BRST
invariant (here standard) gauge fixing with the local action according to the
BFV formalism.

Therefore, we see that the auxiliary BF field $\rho$ is exactly the
well-known St\"uckelberg scalar in Eq. (69).

\vspace{1cm}
\begin{center}
\subsection {\bf Nonlocal Effective Action}
\end{center}

Although in the previous subsection we have performed the integration over
${\cal P}$ and $\overline{\cal P}$ but not over $\overline{\cal C}$ and
${\cal C}$, it is not impossible to consider the opposite procedure, i.e.,
the integration over $\overline{\cal C}$ and ${\cal C}$ but not over
${\cal P}$ and $\overline{\cal P}$ which is dual to the previous integration.
To this end, we consider the BFV formalism in the previous subsection
up to the point where
the integration over the momentum $\pi_{\rho}$ was performed and
the following effective action was obtained:

\begin{eqnarray}
S_{eff} &=& \int\!d^4x~[ -\frac{1}{4}F_{\mu \nu}F^{\mu \nu}
            + \frac{1}{2} m^{2}
               ( A_{\mu} + \partial_\mu \rho )^2 \nonumber \\
 && - A^{\mu}\partial_{\mu}B + \frac{1}{2}\alpha (B)^2
             - \partial_{i}{\overline{\cal C}}\partial^{i}{\cal C}
       + {\overline{\cal P}}\dot{\cal C}
       + {\overline{\cal C}}\dot{\cal P}
       - {\overline{\cal P}}{\cal P}~].
\end{eqnarray}
This action is invariant under the BRST transformation
which have the form
\begin{eqnarray}
\delta_{B}A_{0} &=& -\lambda{\cal P},
~~~~~~\delta_{B}A_{i} = -\lambda\partial_{i}{\cal C},
~~~~~~\delta_{B} \rho = \lambda {\cal C},  \nonumber \\
 \delta_{B}{\cal C}  &=& 0,
          ~~~~~~\delta_{B}{\overline{\cal C}} = - \lambda B,
          ~~~~~~\delta_{B}B = 0, \nonumber \\  
 \delta_{B}{\cal P}  &=& 0,
          ~~~~~~\delta_{B}{\overline{\cal P}} = - \lambda (
          -\partial_{i}F^{0i} + m^{2} (\dot{\rho} + A^{0})) .
\end{eqnarray}

Now, let us perform the integration over ${\cal C}$, ${\overline{\cal C}}$
instead of their conjugated ones ${\cal P}$, ${\overline {\cal P}}$.
First, performing the
integration over the ghost field ${\cal C}$, we get the following
delta function
\begin{equation}
\delta( \partial_{i}\partial^{i}{\overline{\cal C}} 
        -\dot{{\overline{\cal P}}} )~=~\mbox{det}(\partial_{i}\partial^{i})
  \delta( {\overline{\cal C}}-\frac{1}{\partial_{i}\partial^{i}}
  \dot{{\overline{\cal P}}}).    
\end{equation}
Next, performing the integration over ${\overline{\cal C}}$,
we get the unusual non-local form of the effective action
\begin{equation}
S_{eff} = \int\!d^4x~[ -\frac{1}{4}F_{\mu \nu}F^{\mu \nu}
            + \frac{1}{2} m^{2}
               ( A_{\mu} + \partial_\mu \rho )^2
 - A^{\mu}\partial_{\mu}B + \frac{1}{2}\alpha (B)^2
+ \dot{{\overline{\cal P}}}
   \frac{1}{\partial_{i}\partial^{i}}
   \dot{{\cal P}} -{\overline{\cal P}}{\cal P} ]
\end{equation}
which is non-covariant.
Notice that the appearance of the nonlocal term in the ghost action was
a result of the unusual integration. But, we may find
that this form is also obtained by the change of variables
\begin{equation}    
{\cal C} ~\to~ \frac{1}{\partial_{i}\partial^{i}}{\overline{\cal P}} ,~~    
{\overline{\cal C}} ~\to~{\cal P}
\end{equation}
in the Eq. (69). Under these replacements, we have the nonlocal BRST charge
$Q'$ given by
\begin{equation}    
Q'~=~\int\!d^3x~[B{\overline{\cal C}}+(
-\partial_{i}F^{0i} + m^{2} (\dot{\rho} + A^{0}))
 \frac{1}{\partial_{i}\partial^{i}}{\overline{\cal P}} ].    
\end{equation}
Then, the effective action is invariant under the following
nonstandard BRST transformation as
\begin{eqnarray}
\delta_{B}'A_{\mu} &=& -\lambda\partial_{\mu}
                    (\frac{1}{\partial_{i}\partial^{i}}{\overline{\cal P}}) ,
 ~~~~~~\delta_{B}' \rho = \lambda \frac{1}{\partial_{i}\partial^{i}}
                    {\overline{\cal P}} ,  \nonumber \\
 \delta_{B}'{\overline{\cal P}}  &=& 0,
          ~~~~~~\delta_{B}'{\cal P} = - \lambda B,
          ~~~~~~\delta_{B}'B = 0,
\end{eqnarray}
which is non-local and non-covariant and hence can be categorized as the
symmetry recently proposed in QED by Lavelle and McMullan. $^{23}$
Moreover, this nonlocal BRST symmetry yields a conserved current through
the Noether's theorem as follows
\begin{equation}    
J_{B \mu}'~=~F_{\mu \nu}\partial^{\nu} 
           \frac{1}{\partial_{i}\partial^{i}}{\overline{\cal P}}+ 
           m^{2} ( A_{\mu} + \partial_\mu \rho )
           \frac{1}{\partial_{i}\partial^{i}}{\overline{\cal P}} +
           B \partial_{\mu}
           \frac{1}{\partial_{i}\partial^{i}}{\overline{\cal P}}.
\end{equation}
This completes the procedure of the BRST invariant (here nonstandard)
gauge fixing
with the non-local action according to BFV formalism.
As a result, we have recognized that these nonlocal symmetry and the 
conserved current are nothing but
the original local theory performing the change of variable (75).

\vspace{1cm}

\begin{center}
\section{\bf Conclusion}
\end{center}

In conclusion, we have applied the BFT and the BFV method to
covariantly quantize
the second class constraint system of the Abelian Proca model without 
spoiling the unitarity.
First, by applying the BFT method, we have systematically converted
the second class system of the model into the effectively first class
one in the extended phase
space. We have shown the relation that, due to the linear character of the
constraint, the Dirac brackets between the phase space variables in
the original second class system are exactly the Poisson brackets of the
corresponding modified ones in the extended phase space without
$\Phi \rightarrow 0$ limiting procedure of the general formula of BFT $^{7}$
by comparing the Dirac or Faddeev-Jackiw symplectic formalism.

Furthermore, we have noted that, like as this relation, the general relation
(30) of the Dirac brackets in the non-extended phase space and the Poisson
brackets in the extended
phase space is essentially due to the Abelian conversion (31) of the second
class constraint into the first class one and we have added also that some
more general conversion method (35) may be considered without spoiling
this nice relation.

Moreover, we have adopted a new approach, which is more simpler than
the usual
one, in finding the involutive Hamiltonian by using these modified
variables according to the important property (24). Now, with this
first class
constraint we have applied the BFV method to covariantly quantize the
Proca model
without spoiling the unitarity.
By identifying a new  auxiliary field with the
St\"uckelberg scalar we have naturally derived the St\"uckelberg scalar
term related to the explicit gauge symmetry breaking mass term
through the usual local gauge fixing procedure with the standard BRST
symmetry according to the BFV formalism.
We have also analyzed the nonlocal gauge fixing procedure with the
nonstandard BRST symmetry which has been recently
studied by several authors. $^{23}$

As final remarks, we first note  that although we have successfully
applied the
rather simple Abelian case, it is not clear whether the non-Abelian 
generalization of our model is possible or not in a priori.
However, considering the recent failure of the complete
conversion of the second class constraints of this non-Abelian model into the
first class ones, $^{19}$ which does not use the symmetric
$X_{ij}(x,y)$-matrix, and the power of the
formalism with the symmetric $X_{ij}(x,y)$ as noted in Section 2.1 we
expect that
whether we can find the symmetric $X_{ij}(x,y)$ or not is crucial
point for the successful application of
our BFT formalism. We are in progress in this direction. In another
direction,
a new formalism may be considered to solve this problem, but it is not clear
whether the complete conversion of the system into the first class one is
possible and furthermore that
formalism is equivalent to the BFT formalism or not. $^{30}$ On the
other hand, it
is interesting to note that the non-Abelian model of the Chern-Simons theory
allows the symmetric $X_{ij}(x,y)$ and the solution of the first class
system is found by finite iterations. $^{18}$

\vspace{1cm}

\begin{center}
\section*{Appendix A}
\end{center}

In this appendix, we derive the first class Hamiltonians (26) and (27)
in the extended phase space corresponding to the total Hamiltonian $H_{T}$
of (3) and canonical Hamiltonian $H_{c}$ of (4) by using the usual
straightforward approach. $^{13, 17-19, 28}$ Let us first consider the first
class Hamiltonian $\widetilde{H_{T}}$ corresponding to $H_{T}$.
It is given by the infinite series,
\begin{equation}
 \widetilde{H_{T}} = H_{T} + \sum_{n=1}^{\infty} H_{T}^{(n)};
                         ~~~~~H_{T}^{(n)} \sim (\Phi^i)^n,
\end{equation}
satisfying the initial condition,
$\widetilde{H_{T}}(\pi_\mu, A^\mu; 0) = H_{T}$.
The general solution $^{7}$
for the involution of $\widetilde{H_{T}}$ is given by
\begin{equation}
  H_{T}^{(n)} = -\frac{1}{n} \int d^3 x d^3 y d^3 z~
              \Phi^i(x) \omega_{ij}(x,y) X^{jk}(y,z) G_k^{(n-1)}(z)
              ~~~(n \geq 1),
\end{equation}
where the generating functions $G_k^{(n)}$ are given by
\begin{eqnarray}
  G_i^{(0)} &=& \{ \Omega_i^{(0)}, H_{T} \},  \nonumber  \\
  G_i^{(n)} &=& \{ \Omega_i^{(0)}, H_{T}^{(n)} \}_{\cal O}
                    + \{ \Omega_i^{(1)}, H_{T}^{(n-1)} \}_{\cal O}
                                       ~~~ (n \geq 1),
\end{eqnarray}
where the symbol ${\cal O}$ in Eq. (80) represents
that the Poisson brackets are calculated among the original variables, i.e.,
${\cal O}=( A^\mu , \pi_\mu)$.
Here, $\omega_{ij}$ and $X^{ij}$ are the inverse matrices of $\omega^{ij}$
and $X_{ij}$ respectively as in the text. Explicit calculations yields
\begin{eqnarray}
G_1^{(0)} &=& \Omega_2, \nonumber \\
G_2^{(0)} &=& \partial_{i} \partial_i \Omega_{1},
\end{eqnarray}
which are substituted in (80) to obtain $H_{T}^{(1)}$,
\begin{equation}
 H_{T}^{(1)} =  \int d^3x \left[
                  \frac{1}{m}(\partial_{i} \partial_{i} \Phi^1 )\Omega_{1}
                  - \frac{1}{m} \Phi^2 \Omega_{2}
                  \right].
\end{equation}
This is inserted back in Eq. (81) in order to deduce $G_i^{(1)}$ as follows
\begin{eqnarray}
G_1^{(1)} &=& m \Phi^2, \nonumber\\
G_2^{(1)} &=& m \partial_i \partial_i \Phi^1.
\end{eqnarray}
Then, we obtain
$H_{T}^{(2)}$ by substituting $G_i^{(1)}$ in Eq. (76)
\begin{equation}
 H_{T}^{(2)}  =  \int  d^3x \left[
                  - \frac{1}{2} (\partial_i \Phi^1)
                    (\partial_{i} \Phi^1)
                  - \frac{1}{2} (\Phi^2)^2
                  \right].
\end{equation}
Finally, since
\begin{equation}
G_i^{(n)} = 0~~~~~~( n \geq 2),
\end{equation}
due to the proper choice (12) we obtain the complete form of the Hamiltonian
${\widetilde{H}}$ as follows
\begin{equation}
\widetilde H_{T} = H_T + H_{T}^{(1)} + H_{T}^{(2)},
\end{equation}
which, by construction, is strongly involutive,
\begin{equation}
\{\widetilde{\Omega}_i, \widetilde H_{T} \} = 0.
\end{equation}

Similarly, for the canonical Hamiltonian we can easily obtain it's
first class
Hamiltonian $\widetilde{H_{c}}$ as follows
\begin{eqnarray}
\widetilde{H_{c}}=H_{c}+H_{c}^{(1)}+H_{c}^{(2)},
\end{eqnarray}
where
\begin{eqnarray}
&& H_{c}^{(1)} =  \int d^3x \left[
                  m \Phi^1 (\partial_{i} A^{i})
                  - \frac{1}{m}  \Phi^2 \Omega_{2}
                  \right], \\
&& H_{c}^{(2)}  =  \int  d^3x \left[
                   \frac{1}{2} (\partial_i \Phi^1)
                    (\partial_{i} \Phi^1)
                  - \frac{1}{2} (\Phi^2)^2
                  \right].
\end{eqnarray}
Here, we used
\begin{eqnarray}
G_1^{(0)} &=& \Omega_2, \nonumber \\
G_2^{(0)} &=& m^{2} \partial_i A^{i}, \\
G_1^{(1)} &=& m \Phi^2, \nonumber\\
G_2^{(1)} &=& -m \partial_i \partial_i \Phi^1. \nonumber \\
G_{i}^{(n)} &=&0 ~~~~~~(n \geq 2). \nonumber
\end{eqnarray}
Note the differences in $G^{(0)}_{2}$, $G^{(1)}_{2}$ and $\Phi^{1}$-dependent
terms in $H^{(1)}$ and $H^{(2)}$ for the total and canonical Hamiltonians.
Moreover, $\widetilde{H_{c}}$ is also, by construction, strongly involutive,
\begin{equation}
\{\widetilde{\Omega}_i, \widetilde H_{c} \} = 0.
\end{equation}

\vspace{1cm}

\begin{center}
\section*{Appendix B}
\end{center}

In this appendix, we obtain the FJ symplectic brackets
comparing them with both the orthodox Dirac brackets and the modified
Poisson brackets (36) in the extended phase space in the section 2.2.

According to the FJ formalism, $^{22}$
which is to be regarded as the improved version of the Dirac one,
we rewrite the first order
Lagrangian corresponding to the Proca model (1) as
\begin{equation}
{\cal L}^{(0)} = \pi_i \dot{A}^i - {\cal H}^{(0)},
\end{equation}
where the conjugate momenta (2) of the gauge fields and the canonical
Hamiltonian ${\cal H}_c$ in which we denote it as ${\cal H}^{(0)}$
showing the iterative nature of the formalism are used.

In order to find the FJ symplectic brackets
we introduce the sets of the symplectic variables $\xi^{(0)k}$ and
the conjugate momenta $a^{(0)}_k$ as follows
\begin{eqnarray}
\xi^{(0)k} &=& (A^0, A^i, \pi_i), \nonumber \\
a^{(0)}_k &=& (0, \pi_i, 0),
\end{eqnarray}
which are usually read off from the form of the canonical sector of
the first order Lagrangian (94), respectively.

Then, the dynamics of the model is governed by the
invertible symplectic two-form matrix such that
\begin{equation}
f^{(0)}_{kl} (x,y) = \frac{\partial a^{(0)}_l(y)}
                         {\partial\xi^{(0)k}(x)} -
                     \frac{\partial a^{(0)}_k(x)}{\partial \xi^{(0)l}(y)},
\end{equation}
through the equations of motion
\begin{equation}
\dot{\xi}^k (x) =   \int d^3y f^{(0)kl} (x,y)
              \frac{\partial {\cal H}^{(0)}(x)}{\partial \xi^{(0)k}(y)},
\end{equation}
where $f^{(0)kl}(x,y)$ is an inverse of $f^{(0)}_{kl}(x,y)$.
However, in the Proca model the symplectic two-form matrix is given by
\begin{equation}
  f_{kl}^{(0)}(x,y) =
   \left( \begin{array}{ccc}
         0 &          0            &           0    \\
         0 &          0            &           -\delta_{ij}    \\
         0 &          \delta_{ij}            &           0
           \end{array}
    \right)
  \delta^3(x-y)
\end{equation}
showing the matrix $f^{(0)}_{kl}(x,y)$ is singular.
As it happens, the symplectic two-form matrix has a zero mode, i.e.,
$\widetilde{\nu}^{(0)}_l=(v_1, 0, 0)$, where $v_1$ is an arbitrary function.
Furthermore, this zero mode generates constraint
$\Omega^{(1)}$ in the context of the FJ formalism $^{22}$
as follows
\begin{eqnarray}
0 &=& \int d^3x~ v_1(x) \frac{\delta}{\delta A^0(x)}
               \int d^3y {\cal H}^{(0)}(y) \nonumber \\
  &=& - \int d^3x~ v_1 (\partial^i \pi_i + m^2 A^0) \nonumber \\
  &\equiv& - \int d^3x~ v_1 \Omega^{(1)},
\end{eqnarray}
which will be added into the canonical sector of the Lagrangian (94)
enlarging the symplectic phase space with the Lagrange multiplier $\alpha$.
Then, the iterated, first order Lagrangian is given by
\begin{equation}
{\cal L}^{(1)} = \pi_i \dot{A}^i + (\partial^i \pi_i + m^2 A^0) \dot{\alpha}
                - {\cal H}^{(1)},
\end{equation}
where the corresponding first iterated
Hamiltonian ${\cal H}^{(1)}$ is given by
\begin{equation}
{\cal H}^{(1)}(\xi) \mid_{\Omega^{(1)}=0} ~=~
                    \frac{1}{2} (\pi_i)^2
                    + \frac{1}{4}F_{ij}F^{ij}
                    + \frac{1}{2} m^2 (A^0)^2
                    + \frac{1}{2} m^2 (A^i)^2.
\end{equation}

The situations we stand is exactly the same as before except we now
have the first order Lagrangian (100) and the Hamiltonian (101).
In other words,
we can set again the symplectic variables and the conjugate momenta
as follows
\begin{eqnarray}
\xi^{(1)k} &=& (A^0, A^i, \pi_i, \alpha), \nonumber \\
a^{(1)}_k &=& (0, \pi_i, 0, \partial^i \pi_i + m^2 A^0)
\end{eqnarray}
reading off from the Lagrangian (100). From this set of the variables,
the first iterated symplectic two-form matrix is given by
\begin{equation}
  f_{kl}^{(1)}(x,y) =
   \left( \begin{array}{cccc}
         0 &          0            &           0     &  m^2 \\
         0 &          0            &           -\delta_{ij}    &  0 \\
         0 &      \delta_{ij}    &           0     &  -\partial^1_x \\
         -m^2 &  0 & \partial^1_y & 0
           \end{array}
    \right)
  \delta^3(x-y),
\end{equation}
and its inverse matrix is easily obtained
\begin{equation}
  f^{kl (1)} (x,y) =
   \left( \begin{array}{cccc}
    0 &          \frac{1}{m^2} \partial^i_x  &   0     &  -\frac{1}{m^2} \\
    \frac{1}{m^2} \partial^i_x &   0   &  \delta_{ij}    &  0 \\
         0 &          -\delta_{ij}       &   0   &  0 \\
         \frac{1}{m^2} &  0 & 0 & 0
           \end{array}
    \right)
  \delta^3(x-y).
\end{equation}
Now, according to the FJ formalism, this inverse symplectic
two-form matrix gives the symplective brackets of the Proca model
\begin{equation}
\{\xi^k(x), \xi^l(y) \}_{symp} = f^{kl (1)} (x,y)
\end{equation}
in the case of having the invertible symplectic matrix, i.e.,
at the final stage of iteration, such that
\begin{eqnarray}
&&\{ A^0 (x), A^j (y) \}_{symp}=\frac{1}{m^{2}} \partial_{x}^{j}
\delta(x-y),  \nonumber \\
&&\{ A^0 (x), A^0 (y) \}_{symp}=0,   \nonumber  \\
&&\{ A^j (x), A^k (y) \}_{symp}=0,  \nonumber \\
&&\{ \Pi_\mu (x), \Pi_\nu (y) \}_{symp}=0,  \nonumber \\
&&\{ A^i (x), \Pi_j (y) \}_{symp}=\delta_{ij} \delta (x-y),  \nonumber \\
&&\{ A^0 (x), \Pi_\nu (y) \}_{symp}=0, \nonumber \\
&&\{ A^i (x), \Pi_0 (y) \}_{symp}=0
\end{eqnarray}
showing that the symplectic brackets are exactly same both
as the Dirac brackets
and the modified Poisson brackets in the extended phase space (36).

\vspace{1cm}

\begin{center}
\section*{Acknowledgements}
\end{center}

We would like to thank Prof. W. T. Kim for helpful discussions.
The present study was supported by
the Basic Science Research Institute Program,
Ministry of Education, Project No. 95-2414.

\newpage
\vspace{1cm}

\begin{center}
\section*{References}
\end{center}
\begin{description}

\item{1.} P. A. M. Dirac, {\it Lectures on quantum mechanics}
         ( Belfer graduate School,
            Yeshiba University Press, New York, 1964 ).
\item{2.} E. S. Fradkin and G. A. Vilkovisky,
            {\it Phys. Lett.} {\bf B55}, 224 (1975).
\item{3.} M. Henneaux, {\it Phys. Rep.} {\bf C126}, 1 (1985).
\item{4.} C. Becci, A. Rouet and R. Stora,
           {\it Ann. Phys. (N.Y.) } {\bf 98}, 287 (1976);
           I. V. Tyutin, Lebedev Preprint 39 (1975).
\item{5.} T. Kugo and I. Ojima,
           {\it Prog. Theor. Phys. Suppl.} {\bf 66}, 1 (1979).
\item{6.} I. A. Batalin and E. S. Fradkin,
           {\it Nucl. Phys.} {\bf B279}, 514 (1987);
           {\it Phys. Lett.} {\bf B180}, 157 (1986).
\item{7.} I. A. Batalin and I. V. Tyutin,
           {\it Int. J. Mod. Phys. }{\bf A6}, 3255 (1991); E. S. Fradkin,
           {\it Lecture of the Dirac Medal of ICTP 1988 } (Miramare-Trieste
           , 1988)
\item{8.} T. Fujiwara, Y. Igarashi and J. Kubo,
            {\it Nucl. Phys.} {\bf B341}, 695 (1990);
            {\it Phys. Lett.} {\bf B251}, 427 (1990); J. Feinberg and 
 M. Moshe,
           {\it Ann. Phys.} {\bf 206}, 272 (1991).
\item{9.} Y.-W. Kim, S.-K. Kim, W. T. Kim, Y.-J. Park,
           K.Y. Kim, and Y. Kim, {\it Phys. Rev.} {\bf D46}, 4574 (1992).
\item{10.} R. Banerjee, H. J. Rothe and K. D. Rothe,
           {\it Phys. Rev. }{\bf D49}, 5438 (1994).
\item{11.} L. D. Faddeev and S. L. Shatashivili,
           {\it Phys. Lett. }{\bf B167}, 225 (1986);
           O. Babelon, F. A. Shaposnik and C. M. Vialett,
           {\it Phys. Lett. }{\bf B177}, 385 (1986);
           K. Harada and I. Tsutsui, {\it Phys. Lett. }{\bf B183}, 311
(1987).
\item{12.} J. Wess and B. Zumino, {\it Phys. Lett.} {\bf B37}, 95 (1971).
\item{13.} R. Banerjee, {\it Phys. Rev.} {\bf D48}, R5467 (1993).
\item{14.} R. Jackiw, in {\it Topological Investigations of
           Quantized Gauge Theories}, edited by S. Treiman,
           R. Jackiw, B. Zumino and E. Witten
           (World Scientific, Singapore, 1985).
\item{15.} G. Semenoff, {\it Phys. Rev. Lett.} {\bf 61}, 517 (1988);
           G. Semenoff and P. Sodano, {\it Nucl. Phys.} {\bf B328}, 
753 (1989).
\item{16.} R. Banerjee, {\it Phys. Rev. Lett.} {\bf 69}, 17 (1992);
           {\it Phys. Rev.} {\bf D48}, 2905 (1993);
           R. Banerjee, A. Chatterjee and V. V. Sreedhar,
           {\it Ann. Phys. (N.Y.)} {\bf 122}, 254 (1993).
\item{17.} Y.-W. Kim, Y.-J Park, K. Y. Kim and Y. Kim, {\it Phys. Rev.}
           {\bf D51}, 2943 (1995);
           E.-B. Park, Y.-W. Kim, Y.-J Park, Y. Kim, and W. T. Kim,
           {\it Mod. Phys. Lett.} {\bf A10}, 1119 (1995).
\item{18.} W. T. Kim and Y. -J. Park, {\it Phys. Lett.} {\bf B336}, 
376 (1994).
\item{19.} N. Banerjee, R. Banerjee and S. Ghosh, {\it Ann. Phys.}
           {\bf 241}, 237 (1995).
\item{20.} P. A. M. Dirac, {\it Can. J. Math.} {\bf 3}, 1 (1950);
  E. C. G. St\"uckelberg, {\it Helv. Phys. Act. }{\bf 30}, 209 (1957);
  P. Senjanovic,
  {\it Ann. Phys. (N.Y.)}, {\bf 100}, 227 (1976); {\bf 209}, 248(E) (1991).
\item{21.} E. C. G. St\"uckelberg in Ref. 20;
           L. D. Faddeev, {\it Theor. Math. Phys.} {\bf 1}, 1(1970).
\item{22.} Faddeev and R. Jackiw, {\it Phys. Rev. Lett.} {\bf 60}, 
1692 (1988);
J. Barcelos-Neto and C. Wotzasek, {\it Int. J. Mod. Phys.} {\bf A7}, 
4781 (1992);
Y. W. Kim, Y.-J. Park, and Y. Kim, {\it J. Korean Phys. Soc.} {\bf
28}, 773 (1995).
\item{23.} M. Lavelle and David McMullan,
            {\it Phys. Rev. Lett.} {\bf 71}, 3758 (1993); Z. Tang and 
D. Finkelstein,
           {\it Phys. Rev. Lett.} {\bf 73}, 3055 (1944);
            S. J. Rabello and P. Gaete, {\it Phys. Rev.} {\bf D52}, 
7205 (1995);
            H. Shin, Y.-J. Park, Y. Kim, and W. T. Kim,
            {\it J. Korean Phys. Soc.} {\bf 29}, 392 (1996).
\item{24.} N. Banerjee, S. Ghosh and R. Banerjee,
           {\it Nucl. Phys.} {\bf B417}, 257 (1994);
           {\it Phys. Rev.} {\bf D49}, 1996 (1994); R. Amorim and J. B. Neto,
           {\it Phys. Lett.} {\bf B 333}, 413 (1994).
\item{25.} R. Amorim and A. Das, {\it Mod. Phys. Lett.} {\bf A9}, 3453

(1994);
              R. Amorim, {\it Z. Phys.} {\bf C67}, 695 (1995);
              N. Banerjee and R. Banerjee, {\it Mod. Phys. Lett.} {\bf
A11}, 1919 (1996).
\item{26.} T. Maskawa and H. Nakajima, {\it Prog. Theor. Phys.} {\bf
56}, 1295 (1976);
     S. Weinberg, {\it The Quantum Theory of Fields } (Cambridge
University Press, New York, 1995).
\item{27.} L. D. Faddeev and V. N. Popov, {\it Phys. Lett.} {\bf B25},
29 (1967).
\item{28.} J.-H. Cha, Y.-W. Kim, Y.-J. Park, Y. Kim, S.-K. Kim,
           and W. T. Kim, {\it Z. Phys.} {\bf C 69}, 175 (1995).
\item{29.} G. Parisi and N. Sourlas, {\it Phys. Rev. Lett.} {\bf 43},
774 (1979);
     H. Aratya, R. Ingermanson, and A. J. Niemi, {\it Nucl. Phys.} 
{\bf B 307}, 157 (1988);
     A. J. Niemi, {\it Phys. Rev.} {\bf D 36}, 3731 (1987).
\item{30.} C. Bizdadea and S. O. Saliu, {\it Nucl. Phys.} {\bf B456}, 
473 (1995);
{\it ibid.} {\bf B469}, 302 (1996).
\end{description}
\end{document}